\documentclass[aps,prd,reprint,notitlepage,groupedaddress,nofootinbib,preprintnumbers]{revtex4-1}
\usepackage[utf8]{inputenc}
\usepackage{amsmath,amssymb,amsfonts}
\usepackage{graphicx}
\usepackage{color}
\usepackage[colorlinks=true,linkcolor=blue,urlcolor=blue,citecolor=blue]{hyperref}
\usepackage{slashed}
\usepackage{lipsum}
\usepackage{tgtermes}

\def\be{\begin{equation}}
\def\ee{\end{equation}}

\def\bea{\begin{eqnarray}}
\def\eea{\end{eqnarray}}

\def\vec[#1]{\boldsymbol{#1}}
\def\vecs[#1,#2]{\boldsymbol{{#1}_{#2}}}

\def\mes[#1]{d^{3}{#1}}

\def\del{\partial}
\newcommand{\half}{\frac{1}{2}}

\newcommand{\bph}{\bar{\ph}}
\newcommand{\bps}{\bar{\psi}}

\newcommand{\beq}{\begin{equation}}
\newcommand{\eeq}{\end{equation}}
\newcommand{\bc}{}
\newcommand{\nn}{\nonumber}

\newcommand{\mN}{\mathcal{N}}

\newcommand{\f}{\frac}

\newcommand{\al}{\alpha}
\newcommand{\ga}{\gamma}         \newcommand{\Ga}{\Gamma}
\newcommand{\de}{\delta}        
\newcommand{\ep}{\epsilon}

\newcommand{\si}{\sigma}

\newcommand{\ph}{\phi}          \newcommand{\Ph}{\Phi}
\newcommand{\ps}{\psi}

\newcommand{\mO}{\mathcal{O}}

\begin{document}

\title{\boldmath Mass-deformed $\mathcal{N}=3$ Supersymmetric Chern-Simons-Matter Theory}

\author{Karthik Inbasekar}
\email{inbaseka@post.bgu.ac.il}
\affiliation{Department of Physics, Ben-Gurion University of the Negev, Beer-Sheva 84105, Israel, }
\affiliation{Faculty of Exact Sciences, School of Physics and Astronomy, Tel Aviv University, Tel Aviv 69978, Israel}

\author{Lavneet Janagal}
\email{lavneet@theory.tifr.res.in}
\affiliation{Department of Theoretical Physics, Tata Institute of Fundamental Research, 1 Homi Bhabha Road, Mumbai, India 400005}

\author{Ashish Shukla}
\email{ashish@uvic.ca}
\affiliation{Department of Physics and Astronomy, University of Victoria, 3800 Finnerty Road, Victoria, BC, Canada V8P 5C2}

\preprint{\texttt{TIFR/TH/19-29}}

\begin{abstract}
\noindent
The maximal extension of supersymmetric Chern-Simons theory coupled to fundamental matter has $\mathcal{N} = 3$ supersymmetry.  In this short note, we provide the explicit form of the action for the mass-deformed $\mathcal{N} = 3$ supersymmetric $U(N)$ Chern-Simons-Matter theory. The theory admits a unique triplet mass deformation term consistent with supersymmetry. We explicitly construct the mass-deformed $\mN=3$ theory in $\mN=1$ superspace using a fundamental and an anti-fundamental superfield. 
\end{abstract}

\maketitle


\section{Introduction}
\label{Intro}
\noindent
Pure Chern-Simons theory is topological in nature and has no propagating degrees of freedom. When coupled to matter, the Chern-Simons gauge field attaches magnetic fluxes to the interacting matter quanta that gives rise to interesting physical phenomena such as the Aharonov-Bohm effect \cite{Aharonov:1959fk}.  Chern-Simons theories coupled to matter are applicable in diverse physical systems from quantum Hall effect  \cite{10.1007/BFb0113369, XGW, Witten:2015aoa, Tong:2016kpv}  in condensed matter theory to quantum gravity in high energy physics \cite{Witten:1988hc}.
 
Non-abelian Chern-Simons theories coupled to matter in the fundamental representation of $U(N)$ or $SU(N)$ are exactly solvable in the \lq t Hooft large $N$ limit. The theories enjoy a strong-weak bosonization duality \cite{Giombi:2011kc,Aharony:2011jz,Maldacena:2011jn,Maldacena:2012sf,Aharony:2012nh} that has been rigorously tested in the planar limit over the past decade. A simple example of the duality is between a Chern-Simons theory coupled to a fundamental boson in the Wilson-Fisher limit and a Chern-Simons theory coupled to a fundamental fermion.  Under a mapping of the parameters in the large $N$ limit, the physical observables in one theory map to the other. 

The evidence for the duality consists of several rigorous large $N$ computations such as thermal partition functions \cite{Aharony:2012ns,Yokoyama:2013pxa,Takimi:2013zca,Jain:2013py}, correlation functions \cite{Aharony:2011jz,Aharony:2012nh,GurAri:2012is,Bedhotiya:2015uga,Gur-Ari:2015pca,Geracie:2015drf,Turiaci:2018dht,Yacoby:2018yvy,Inbasekar:2019wdw}, and $S$-matrices \cite{Jain:2014nza,Dandekar:2014era,Inbasekar:2015tsa,Yokoyama:2016sbx} to name a few. Significant evidence has accumulated over the years that the duality holds true along RG flows \cite{Minwalla:2015sca,Aharony:2018pjn}. These RG flows relate the bosonization duality to dualities in supersymmetric Chern-Simons-Matter theories \cite{Jain:2012qi,Jain:2013gza,Aharony:2019mbc}. In this context, the bosonization duality generalizes the already well known dualities such as Giveon-Kutasov duality in supersymmetric Chern-Simons-Matter theories \cite{Benini:2011mf,Park:2013wta,Aharony:2013dha}. For other recent works in the field, we refer the reader to \cite{Choudhury:2018iwf,Dey:2018ykx,Dey:2019ihe,Gur-Ari:2016xff,Halder:2019foo} and in particular to the applications of the finite $N$ generalizations of the duality \cite{Radicevic:2015yla,Aharony:2015mjs,Seiberg:2016gmd,Karch:2016sxi,Hsin:2016blu,Cordova:2017kue,Metlitski_2017,Cordova:2018qvg}.

In addition to being testing grounds for the duality, amplitudes in Chern-Simons-Matter theories can be used to demonstrate anyonic effects in these theories. In particular, the $S$-matrices in these theories display non-trivial phenomena arising from the anyonic nature of matter. For instance, unitarity of the scattering amplitudes in these theories requires a modification of the rules of crossing symmetry \cite{Jain:2014nza,Inbasekar:2015tsa,Yokoyama:2016sbx}. Furthermore, amplitudes in supersymmetric Chern-Simons-Matter theories (especially with a high degree of supersymmetry) have unique properties. In \cite{Inbasekar:2015tsa}, exact $2\to2$ amplitudes were computed to all orders in the 't Hooft coupling for $\mN=2$ supersymmetric Chern-Simons-Matter theory. These amplitudes enjoy remarkable symmetry properties, such as dual superconformal symmetry \cite{Inbasekar:2017sqp} and Yangian symmetry \cite{Yangian} exact to all loops. Furthermore, using recursion relations, arbitrary $n$-point tree level amplitudes have also been constructed \cite{Inbasekar:2017ieo}. These symmetry properties are perhaps a general feature of supersymmetric Chern-Simons-Matter theories with $\mN\geq 2$ supersymmetry. It is for instance well known that in $\mathcal{N}=6$ supersymmetric Chern-Simons-Matter theory n-point amplitudes enjoy invariance under dual superconformal symmetry \cite{Gang:2010gy}.

Thus, it is interesting to study scattering amplitudes in supersymmetric Chern-Simons-Matter theories with higher supersymmetry. The technical route towards such computations is by using the superspace methodology. Since an explicit superspace formalism does not exist beyond $\mN > 2$ superspace, one can formulate higher supersymmetric theories in $\mN=1$ or $\mN=2$ superspace. The technology for computing the dynamics via Dyson-Schwinger equations in $\mN=1$ superspace is already well established \cite{Inbasekar:2015tsa}. In this work, we initiate a program to set up higher supersymmetric theories in $\mN=1$ superspace, with a future goal to study scattering amplitudes in these theories.

The maximal supersymmetric extension of $U(N)$ Chern-Simons theory coupled to fundamental matter has $\mathcal{N} = 3$ supersymmetry \cite{Kao:1995gf, Kapustin:1999ha, Gaiotto:2007qi, Chang:2012kt}. Though action for the $\mathcal{N} = 3$ theory without a mass term is known and a general action exists in $\mN=2$ superspace, the explicit form of the mass-deformed action and the supersymmetry transformations have not been written down in the literature. In this paper, we present the explicit form of the mass-deformed action for the $\mathcal{N} = 3$ supersymmetric Chern-Simons theory coupled to fundamental matter and write it in $\mN=1$ superspace.  Our results fill in a gap in the existing literature. In a follow up paper \cite{InbasekarN3}, we explore the Dyson-Schwinger method for computation of exact $S$-matrices in the mass-deformed $\mathcal{N} = 3$ theory. Several computations, such as the ones described in the above paragraphs (\cite{Inbasekar:2017ieo,Inbasekar:2017sqp,Inbasekar:2019wdw}), should also be doable using the Lagrangian eq. \eqref{n3sup} derived in this paper.

The paper is organized as follows. In section \ref{mdn3} we study mass-deformations of the free $\mathcal{N} = 3$ Wess-Zumino model, which serves as a warm up exercise. The model exhibits two consistent mass-deformation terms, transforming as a singlet or a triplet under the $R$-symmetry group $SO(3)_R$. In section \ref{mainsec} we discuss the mass-deformed $\mathcal{N} = 3$ $U(N)$ Chern-Simons-Matter theory. For completeness, we first write down the action for the superconformal theory as well as the supersymmetry transformations which leave the action invariant, and then mass-deform the action. The theory admits a unique mass deformation term, transforming as a triplet under the $R$-symmetry group, in agreement with the argument in \cite{Cordova:2016xhm}. We then repackage the mass-deformed action in the language of $\mathcal{N} = 1$ superfields. We end with concluding comments in section \ref{discuss}. Appendix \ref{conventions} sets the notations and conventions we use. Appendices \ref{massdeform}-\ref{varn3l} provide additional supplementary material. 
\vspace{-3mm}
\section{Mass-deformed $\mathcal{N} = 3$ Wess-Zumino Model}
\label{mdn3}
In this section,  as a warm up exercise, we discuss mass deformations consistent with supersymmetry in the free $\mathcal{N} = 3$ Wess-Zumino model. This section sets the notation for the rest of the discussion (for some details regarding the conventions we follow see appendix \ref{conventions}). In $2+1$ dimensions a theory with $\mN$ supersymmetries has the $R$-symmetry group $SO(\mN)_R$ (the subscript $R$ denotes $R$-symmetry). Consequently, for the three dimensional theories of interest to us the $R$-symmetry group is $SO(3)_R$. The matter content of the $\mathcal{N} = 3$ Wess-Zumino model consists of bosonic and fermionic $SO(3)_R\sim SU(2)_R$ doublets $\phi_A, \psi_{\alpha A}$. Here $A, B,\ldots$ are $SU(2)_R$ indices and take the values $1,2$, and $\alpha, \beta$ denote spinor indices, also taking values $1,2$.  The $\mN=3$ supercharge $Q_{AB\alpha}$ transforms as a vector under $SU(2)_R$, and is symmetric under the interchange of the $R$-symmetry indices. The action for the free mass-deformed theory is given by
\be
\begin{split}
\label{actn3}
S_{WZ}^{{free}} = \int d^3x \,\Big[&-\del^\mu \bar\phi^A \del_\mu \phi_A - \bar\phi^A M_A^{~\,D} M_D^{~\,E} \phi_E \\
&+ \bar\psi^{\alpha A} i \del_{\alpha}^{~\,\beta} \psi_{\beta A} + \bar\psi^{\alpha A} M_A^{~\,D} \psi_{\alpha D} \Big].
\end{split}
\ee
For a three-vector $V_\mu$, we have the symmetric spinor representation
$V_{\alpha\beta} = V_\mu (\gamma^\mu)_{\alpha\beta}.$ The mass matrix $M_A^{~\,B}$ in the action eq.\ \eqref{actn3} is the most general mass deformation consistent with the symmetries of the theory. As it turns out, the free theory admits two consistent mass-deformations, $M_A^{~\,B} = m \delta_A^{~\,B}$ or $M_A^{~\,B} = m (\sigma_3)_A^{~\,B}$, with $(\sigma_i)_A^{~\,B}, i = 1,2, 3$ denoting the standard Pauli matrices, as we show below.

The most general supersymmetry transformations for the free $\mathcal{N} = 3$ Wess-Zumino model follow from Lorentz invariance and are given by
\begin{equation}
\begin{split}
\label{gensusy3}
&Q_{BC\alpha} \phi_A = \chi_1 \, \psi_{\alpha (B} \, \epsilon_{C)A} ,\\
&Q_{BC\alpha} \bar\phi^A = \tilde\chi_1 \, \bar\psi_{\alpha (B} \, \delta_{C)}^{~~A} ,\\
&Q_{BC\alpha} \psi_{\beta A} = \chi_2 \, \del_{\alpha\beta} \phi_{(B} \, \epsilon_{C)A} + \chi_3 \, C_{\alpha\beta} \phi_{(B} M_{C)A} ,\\
&Q_{BC\alpha} \bar\psi^{\beta A} = \tilde\chi_2 \, \del_{\alpha}^{~\,\beta} \bar\phi_{(B} \, \delta_{C)}^{~~A} + \tilde\chi_3 \, \delta_{\alpha}^{~\,\beta} \bar\phi_{(B} M_{C)}^{~~A},
\end{split}
\end{equation}
where $\chi_1, \chi_2$ etc. are undetermined coefficients, and parentheses on indices denote symmetrization. We now demonstrate two ways to mass deform the theory. Supersymmetric invariance of the action eq.\ \eqref{actn3} when acted upon by the the supercharges provides constraints on the unknown parameters in the supersymmetry transformations eq.\ \eqref{gensusy3}.

\subsection{Case 1: $M_A^{~\,B} = m \,\delta_A^{~\,B}$}
In the first case, we consider the mass deformation $M_A^{~\,B} = m\, \delta_A^{~\,B}$ in eqs.\ \eqref{actn3} and \eqref{gensusy3}. A priori, the coefficients $\chi_i, \tilde\chi_i$ in the supersymmetry transformations eq.\eqref{gensusy3} are arbitrary. To relate them, we take the complex conjugates of the first and third supersymmetry transformations in eq.\eqref{gensusy3}, and compare the results with the second and fourth transformations.\footnote{Complex conjugation raises the lower $R$-symmetry indices and vice versa. Thus, $(Q_{AB\alpha})^* = Q^{AB}_{\alpha}$ and $(\epsilon_{AB})^* = \epsilon^{AB}$. Also, for the scalar and spinor fields we have $(\phi_A)^* = \bar\phi^A$ and $(\psi_{\alpha A})^* = \bar\psi^A_\alpha$. See appendix \ref{conventions} for discussion.} This gives us the relations\footnote{The complex conjugate of the LHS of the third SUSY transformation in eq.\eqref{gensusy3} is $(Q_{BC\alpha} \psi_{\beta A})^* = \bar\psi^A_\beta \overleftarrow{Q}^{BC}_\alpha = -\, Q^{BC}_\alpha \bar\psi^A_\beta.$ The additional $-$ sign contribution is important.}
\be
\label{ccrel}
\tilde\chi_1 = - \chi_1^*, \,\, \tilde\chi_2 = \chi_2^*, \,\, \tilde\chi_3 = - \chi_3^*.
\ee

We next impose the condition that the action eq.\eqref{actn3} must be invariant under the supersymmetry transformations eq.\eqref{gensusy3},
\be
\label{invcond1}
Q_{BC\alpha}\left( S^{free}_{WZ} \right)=0,
\ee
up to surface terms. For the present case where $M_A^{~\,B} = m\, \delta_A^{~\,B}$, or equivalently, $M_{AB} = m \,\epsilon_{AB}$, we get (see appendix \ref{massdeform})
\begin{equation}
\begin{split}
\label{cond1}
&\tilde\chi_1 = -i \chi_2, ~\tilde\chi_2 = i \chi_1;\\
&\chi_2 = i\chi_3, ~\tilde\chi_2 = i \tilde\chi_3;\\
&\tilde\chi_1 = \chi_3, ~\tilde\chi_3 = \chi_1.
\end{split}
\end{equation}
Combining the relations above with the relations in eq.\eqref{ccrel}, we get
\begin{equation}
\begin{split}
\label{cond1c}
&\chi_1^{~*} =i \chi_2, ~\chi_2^{~*} = i \chi_1;\\
&\chi_2 = i\chi_3, ~\chi_2^{~*} = - i \chi_3^{~*};\\
&\chi_1^{~*} = - \chi_3, ~\chi_3^{~*} = - \chi_1.
\end{split}
\end{equation}
It is easy to see that these form a consistent set of relations amongst the unknown parameters $\chi_i$. It is thus clear that the mass matrix $M_A^{~\,B} = m\, \delta_A^{~\,B}$ is an allowed mass deformation. With this mass matrix, the action eq.\eqref{actn3} takes the form
\be
\begin{split}
\label{actn3m}
S_{WZ}^{{free}} = \int d^3x\, \Big[&-\del^\mu \bar\phi^A \del_\mu \phi_A - m^2 \bar\phi^A \phi_A \\
&+ \bar\psi^{\alpha A} i \del_{\alpha}^{~\,\beta} \psi_{\beta A} + m \, \bar\psi^{\alpha A} \psi_{\alpha A} \Big].
\end{split}
\ee
Also, the supersymmetry transformations eq.\eqref{gensusy3} become
\begin{equation}
\begin{split}
\label{gsusy3}
&Q_{BC\alpha} \phi_A = \chi_1 \, \psi_{\alpha (B} \, \epsilon_{C)A} ,\\
&Q_{BC\alpha} \bar\phi^A = - \chi_1^{~*} \, \bar\psi_{\alpha (B} \, \delta_{C)}^{~~A} ,\\
&Q_{BC\alpha} \psi_{\beta A} = - \chi_1^{~*} \left[i \del_{\alpha\beta}  +  m C_{\alpha\beta}\right]\phi_{(B} \, \epsilon_{C)A} ,\\
&Q_{BC\alpha} \bar\psi^{\beta A} = \chi_1 \big[ i \del_{\alpha}^{~\,\beta}  + m \delta_{\alpha}^{~\,\beta}  \big]\bar\phi_{(B} \, \delta_{C)}^{~~A}.
\end{split}
\end{equation}
In eq.\eqref{gsusy3}, the parameter $\chi_1$ is still arbitrary.

\subsection{Case 2: $M_A^{~\,B} = m\, (\sigma_3)_A^{~\,B}$ }\label{triplet}
In the second case, we choose the mass deformation $M_A^{~\,B} = m \, (\sigma_3)_A^{~\,B}$. This chooses a specific direction in the $R$-symmetry space, and in fact breaks the $SU(2)_R$  symmetry to $U(1)_R$. Complex conjugation of the first and third supersymmetry transformations in eq.\ \eqref{gensusy3} with the choice $M_A^{~\,B} = m\, (\sigma_3)_A^{~\,B}$ followed by comparison with the second and fourth transformations gives the conditions
\be
\label{ccrela}
\tilde\chi_1 = - \chi_1^*, \,\, \tilde\chi_2 = \chi_2^*, \,\, \tilde\chi_3 = \chi_3^*.
\ee
Also, the supersymmetric invariance of the action eq.\eqref{invcond1} now gives (see appendix \ref{massdeform})
\begin{equation}
\begin{split}
\label{cond2}
&\tilde\chi_1 = -i \chi_2, ~\tilde\chi_2 = i \chi_1;\\
&\chi_2 = - i\chi_3, ~\tilde\chi_2 = i \tilde\chi_3;\\
&\tilde\chi_1 = - \chi_3, ~\tilde\chi_3 = \chi_1.
\end{split}
\end{equation}
Combining these with eq.\eqref{ccrela}, we get the constraints
\begin{equation}
\begin{split}
\label{cond2c}
&\chi_1^{~*} =i \chi_2, ~\chi_2^{~*} = i \chi_1;\\
&\chi_2 = - i\chi_3, ~\chi_2^{~*} = i \chi_3^{~*};\\
&\chi_1^{~*} = \chi_3, ~\chi_3^{~*} = \chi_1.
\end{split}
\end{equation}
As can be clearly seen, the relations in eq.\eqref{cond2c} are also mutually consistent, and hence the choice  $M_A^{~\,B} = m \, (\sigma_3)_A^{~\,B}$ is another possible mass deformation for the free theory. With this choice of the mass deformation term, the action eq.\eqref{actn3} takes the form
\be
\begin{split}
\label{actn3m2}
S_{WZ}^{{free}} = \int &d^3x \, \Big[ -\del^\mu \bar\phi^A \del_\mu \phi_A - m^2 \bar\phi^A \phi_A \\
&+ \bar\psi^{\alpha A} i \del_{\alpha}^{~\,\beta} \psi_{\beta A} + m \bar\psi^{\alpha A} (\sigma_3)_A^{~\,B} \psi_{\alpha B} \Big],
\end{split}
\ee
and the associated supersymmetry transformations are
\begin{equation}
\begin{split}
\label{susy3m2}
&Q_{BC\alpha} \phi_A = \chi_1 \, \psi_{\alpha (B} \, \epsilon_{C)A} ,\\
&Q_{BC\alpha} \bar\phi^A = - \chi_1^* \, \bar\psi_{\alpha (B} \, \delta_{C)}^{~~A} ,\\
&Q_{BC\alpha} \psi_{\beta A} = - i \chi_1^* \, \del_{\alpha\beta} \phi_{(B} \, \epsilon_{C)A} + m \chi_1^* \, C_{\alpha\beta} \phi_{(B} (\sigma_3)_{C)A} ,\\
&Q_{BC\alpha} \bar\psi^{\beta A} = i \chi_1 \, \del_{\alpha}^{~\,\beta} \bar\phi_{(B} \, \delta_{C)}^{~~A} + m \chi_1 \, \delta_{\alpha}^{~\,\beta} \bar\phi_{(B} (\sigma_3)_{C)}^{~~A} .
\end{split}
\end{equation}
Here, $\chi_1$ is an arbitrary complex number. Note the explicit appearance of the $(\si_3)_{A}^{~\,B}$ in the fermion mass term that breaks the $SU(2)_R$ symmetry to $U(1)_R$. Thus the free $\mathcal{N} = 3$ Wess-Zumino model admits two possible consistent mass deformations.

\section{$\mathcal{N} = 3$ Chern-Simons theory coupled to fundamental matter}
\label{mainsec}
\subsection{The Massless Lorentzian Theory}
\label{lorcs3}
We now consider the Lorentzian $\mathcal{N} = 3$ Chern-Simons-Matter theory, with gauge group $U(N)$.  The action is available in $\mN=2$ superspace in \cite{Gaiotto:2007qi}. For our purposes, we make use of the action manifestly written in $SU(2)_R$ notation (see eq. D.18 of \cite{Chang:2012kt}). The action for the $\mN=3$ theory is given by
\begin{equation}
\begin{split}
\label{lo3cs}
&S_{CS} = \int d^3x \bigg[-\frac{\kappa}{4\pi} \epsilon^{\mu\nu\rho} \, \text{Tr}\bigg(A_\mu \del_\nu A_\rho - \frac{2i}{3} A_\mu A_\nu A_\rho\bigg) \\
&\qquad\qquad\qquad\quad+ i \bar\psi^A \slashed{D} \psi_A - D^\mu \bar\phi_A D_\mu \phi^A \\
&- \frac{4\pi^2}{\kappa^2} (\bar\phi_A \phi^B) (\bar\phi_B \phi^C) (\bar\phi_C \phi^A) + \frac{4\pi}{\kappa} (\bar\phi_A \phi^B) (\bar\psi^A \psi_B) \\
&+ \frac{2\pi}{\kappa}(\bar\psi^A \phi_B) (\bar\phi^B \psi_A) -\frac{4\pi}{\kappa} (\bar\psi^A \phi_A)(\bar\phi^B \psi_B) \\
&+ \frac{2\pi}{\kappa} (\bar\psi^A \phi_A) (\bar\psi^B \phi_B) + \frac{2\pi}{\kappa} (\bar\phi^A \psi_A) (\bar\phi^B \psi_B) \bigg],
\end{split}
\end{equation}
where $D_\mu$ is the gauge covariant derivative. The last three lines in the above action contain the interaction terms for the theory. Note that parentheses surrounding two fields denote a contraction of the gauge indices. Once again, by imposing Lorentz invariance and simple dimensional analysis, one can write down the most general supersymmetry transformations for the theory. The invariance of the action when acted upon by the supercharges then fixes the form of the transformations uniquely.  The supersymmetry transformations turn out to be
\begin{equation}
\begin{split}
\label{sun3f}
&Q_{BC\alpha} \phi_A = \chi_1 \, \psi_{\alpha (B} \, \epsilon_{C)A} ,\\
&Q_{BC\alpha} \bar\phi^A = - \chi_1 \, \bar\psi_{\alpha (B} \, \delta_{C)}^{~~A} ,\\
&Q_{BC\alpha} \psi_{\beta A} = -i \chi_1 \, D_{\alpha\beta} \phi_{(B} \, \epsilon_{C)A} \\ &+\frac{2\pi}{\kappa}\chi_1 \, C_{\alpha\beta} (\bar\phi_A \phi_{(B}) \phi_{C)} +\frac{2\pi}{\kappa}\chi_1 \, C_{\alpha\beta} (\bar\phi_{(B} \phi_{C)}) \phi_{A},\\
&Q_{BC\alpha} \bar\psi^{\beta A} = i \chi_1 \, D_{\alpha}^{~\,\beta} \bar\phi_{(B} \, \delta_{C)}^{~~A} \\
&+\frac{2\pi}{\kappa}\chi_1 \, \delta_{\alpha}^{~\,\beta}(\bar\phi_{(B} \phi^{A}) \bar\phi_{C)} - \frac{2\pi}{\kappa}\chi_1 \, \delta_{\alpha}^{~\,\beta} (\bar\phi_{(B} \phi_{C)}) \bar\phi^{A},\\
&Q_{BC\alpha} A^a_\mu = - \,\frac{4\pi}{\kappa}\chi_1 (\gamma_\mu)_\alpha^{~\,\beta} \bar\phi^i_{(B} (T^a)_i^{~j} \psi_{C)\beta j} \\
&- \frac{4\pi}{\kappa}\chi_1 (\gamma_\mu)_\alpha^{~\,\beta} \bar\psi^i_{\beta (B} (T^a)_i^{~j} \phi_{C)j} ,
\end{split}
\end{equation}
where $\chi_1$ is an arbitrary real number. $(T^a)_i^{~\,j}$ are the generators of the gauge group $U(N)$, with $a,b, \cdots$ denoting the gauge indices. See appendix \ref{varn3l} for a derivation of the supersymmetry transformations eq.\ \eqref{sun3f}.  

\subsection{The Mass-deformed Lorentzian Theory}
\label{mdeflt}
We now consider the mass-deformation of the superconformal $\mathcal{N} = 3$ Chern-Simons-Matter theory, with the action given by eq.\ \eqref{lo3cs}. As it turns out, unlike the free Wess-Zumino model, the interacting Chern-Simons theory admits a unique mass deformation term that should transform as a triplet\footnote{We thank Ken Intriligator for helpful communication.} under $SU(2)_R$ \cite{Cordova:2016xhm}. The most general mass deformation term would be of the form $m_0 \mO$, where $\mO$ is an operator of mass dimension 2. Following \S\ref{triplet}, we find that the most general mass-dependent terms that can be added to the action eq.\ \eqref{lo3cs} and still lead to closure  under supersymmetry transformations are given by
\begin{equation}
\begin{split}
\label{lo3csn}
S_{CS}^{\,mass} = \int d^3x \, \bigg[&- m_0^2 (\bar\phi^A \phi_A) + m_0 (\bar\psi^A (\sigma_3)_A^{~~B} \psi_B) \\
&+ \frac{4\pi m_0}{\kappa} (\bar\phi^A \phi_A)  (\bar\phi^C (\sigma_3)_C^{~~D} \phi_D) \bigg].
\end{split}
\end{equation}
The full mass-deformed action, eq.\ \eqref{lo3cs} with the added mass term eq.\ \eqref{lo3csn}, is invariant under the supersymmetry transformations given by
\begin{equation}
\begin{split}
\label{sun3fn}
&Q_{BC\alpha} \phi_A = \chi_1 \, \psi_{\alpha (B} \, \epsilon_{C)A} ,\\
&Q_{BC\alpha} \bar\phi^A = - \chi_1 \, \bar\psi_{\alpha (B} \, \delta_{C)}^{~~A} ,\\
&Q_{BC\alpha} \psi_{\beta A} = -i \chi_1 \, D_{\alpha\beta} \phi_{(B} \, \epsilon_{C)A} +\frac{2\pi}{\kappa}\chi_1 \, C_{\alpha\beta} (\bar\phi_A \phi_{(B}) \phi_{C)}\\ &+\frac{2\pi}{\kappa}\chi_1 \, C_{\alpha\beta} (\bar\phi_{(B} \phi_{C)}) \phi_{A} + m_0 \chi_1 C_{\alpha\beta} (\sigma_3)_{A(B} \phi_{C)}\, ,\\
&Q_{BC\alpha} \bar\psi^{\beta A} = i \chi_1 \, D_{\alpha}^{~\,\beta} \bar\phi_{(B} \, \delta_{C)}^{~~A} +\frac{2\pi}{\kappa}\chi_1 \, \delta_{\alpha}^{~\,\beta}(\bar\phi_{(B} \phi^{A}) \bar\phi_{C)} \\
&- \frac{2\pi}{\kappa}\chi_1 \, \delta_{\alpha}^{~\,\beta} (\bar\phi_{(B} \phi_{C)}) \bar\phi^{A} + m_0 \chi_1 \delta_\alpha^{~\,\beta} (\sigma_3)_{(B}^{~~\,A} \bar\phi_{C)} \, , \\
&Q_{BC\alpha} A^a_\mu = - \,\frac{4\pi}{\kappa}\chi_1 (\gamma_\mu)_\alpha^{~\,\beta} \bar\phi^i_{(B} (T^a)_i^{~j} \psi_{C)\beta j} \\
&- \frac{4\pi}{\kappa}\chi_1 (\gamma_\mu)_\alpha^{~\,\beta} \bar\psi^i_{\beta (B} (T^a)_i^{~j} \phi_{C)j} ,
\end{split}
\end{equation}
with $\chi_1$ being an arbitrary real number. The transformations in eq.\ \eqref{sun3fn} above are a generalization of the ones in eq.\ \eqref{sun3f}, with additional mass-dependent terms required for the invariance of the mass-deformed action.

\subsection{The Mass-deformed Action in $\mathcal{N} = 1$ Superspace}
\label{mdefsup}
In this section, we express the $\mathcal{N} = 3$ mass-deformed action in $\mathcal{N} = 1$ superspace.  We present the Euclidean action, since it is more useful for computations in $\mN=1$ superspace \cite{InbasekarN3}. The Euclidean form is easily obtained from eqs.\ \eqref{lo3cs}, \eqref{lo3csn} by an analytic continuation given by 
\begin{equation}
\begin{split}
\label{analle}
&t_L \rightarrow - i t_E \, , ~ A_0^L \rightarrow i A_2^E,\\
&\psi_L \rightarrow - i \psi_E\, , ~ \bar\psi_L \rightarrow  i \bar\psi_E.
\end{split}
\end{equation}
We pause here to note that one has to be careful with the Euclidean continuation involving spinors. We often
use the definition of the charge conjugation matrix to define
\beq
\bps^\beta_L \equiv (\ps^\dagger)^\al (\ga^0_L)_\al^{\ \beta} \equiv C^{\beta\al}\ps^*_\al\ .
\eeq
From our choice of the Dirac matrices we see that
\beq
(\ps_\al)_L=\begin{bmatrix}
              \ps_1 \\
	      \ps_2
             \end{bmatrix} \ , \
(\bps^\al)_L=i \begin{bmatrix}
              \ps_2^*  & -\ps_1^*
             \end{bmatrix} \ .
\eeq
It is easy to check that this is consistent with the definition of $\bps$ using the charge
conjugation matrix. However, the Euclidean continuation gives
\beq
\bps^\beta_E \equiv (\ps^\dagger)^\al (\ga^2_E)_\al^{\ \beta}= - \begin{bmatrix}
              \ps_2^*  & -\ps_1^*
             \end{bmatrix} \ ,
\eeq
and therefore the Euclidean continuation of $\bps^\beta_L \equiv C^{\beta\al}(\ps_{\al L})^*$ would give identical results provided we use $\ps_L \to -i \ps_E$, as in eq.\ \eqref{analle}. The mass-deformed Euclidean action turns out to be
\begin{equation}
\begin{split}
\label{eu3cs}
&S_{CS}^{\texttt{Euc}} = \int d^3x \, \bigg[\frac{i\kappa}{4\pi} \epsilon^{\mu\nu\rho} \, \text{Tr}\bigg(A_\mu \del_\nu A_\rho - \frac{2i}{3} A_\mu A_\nu A_\rho\bigg) \\
&- i \bar\psi^A \slashed{D} \psi_A + D^\mu \bar\phi_A D_\mu \phi^A + \frac{4\pi^2}{\kappa^2} (\bar\phi_A \phi^B) (\bar\phi_B \phi^C) (\bar\phi_C \phi^A) \\
&- \frac{4\pi}{\kappa} (\bar\phi_A \phi^B) (\bar\psi^A \psi_B) - \frac{2\pi}{\kappa}(\bar\psi^A \phi_B) (\bar\phi^B \psi_A)\\
&+\frac{4\pi}{\kappa} (\bar\psi^A \phi_A)(\bar\phi^B \psi_B) + \frac{2\pi}{\kappa} (\bar\psi^A \phi_A) (\bar\psi^B \phi_B) \\
&+ \frac{2\pi}{\kappa} (\bar\phi^A \psi_A) (\bar\phi^B \psi_B) + m_0^2 (\bar\phi^A \phi_A) - m_0 (\bar\psi^A (\sigma_3)_A^{~~B} \psi_B) \\
&- \frac{4\pi m_0}{\kappa} (\bar\phi^A \phi_A)  (\bar\phi^C (\sigma_3)_C^{~~D} \phi_D) \bigg].
\end{split}
\end{equation}

Now, the $\mN=3$ action in $\mN=1$ superspace can be constructed from a pair of fundamental and anti-fundamental chiral multiplets $(\Phi^+_i,\bar{\Phi}^{-i})$ coupled to a $U(N)$ gauge superfield $\Ga_\al^a (T^a)_i^{\ j}$. We refer the reader to Appendix A of \cite{Inbasekar:2015tsa} for the notation we use for the $\mN=1$ superspace. Since we intend to formulate the $\mN=3$ theory in $\mN=1$ superspace, only the $SO(2)_R$ subgroup of the full $SO(3)_R$ symmetry is manifest. Thus $\Phi^+$ and $\Ph^-$ transform under the two inequivalent one dimensional representations of the residual $SO(2)_R$ symmetry. It is convenient to assign the $SO(2)_R$ charges $(+\f{1}{2},-\f{1}{2})$ for the superfields $(\Ph^+_i,\Phi^-_i)$ respectively. It follows that the complex conjugate fields $(\bar{\Ph}^{+i},\bar{\Ph}^{-i})$ have $SO(2)_R$ charges $(-\f{1}{2},+\f{1}{2})$ respectively. \footnote{A general element of $SO(2)$ is represented by the rotation matrix
\begin{center}
$$
R(\al)=
\begin{pmatrix}
 \cos\al & -\sin\al\\
 \sin\al & \cos\al
\end{pmatrix}
\ \forall \ \al \in (-\pi, \pi].
$$
\end{center}
The ``standard irreps'' of $SO(2)$ are of the form
$$
D^{(n)}(\al) z_n = e^{-i\al n} z_n \ , \ n=0,\pm1,\pm2,\ldots
$$
Baring the trivial representation at $n=0$, for each integer $n$ there exist two inequivalent one dimensional irreps of $SO(2)$.} The $\mathcal{N} = 3$ action written in the language of $\mathcal{N} = 1$ superfields takes the form
\begin{widetext}
\begin{equation}
\begin{split}
\label{n3sup}
\mathcal{S}^{\texttt{Euc}}_{CS} = - \int d^3x \, d^2\theta \, \bigg[ &\frac{\kappa}{4\pi} \, \text{Tr}\bigg(- \frac{1}{4} D^\alpha \Gamma^\beta D_\beta \Gamma_\alpha +\frac{i}{6}\, D^\alpha \Gamma^\beta \{\Gamma_\alpha, \Gamma_\beta \} + \frac{1}{24} \{\Gamma^\alpha, \Gamma^\beta \} \{\Gamma_\alpha, \Gamma_\beta \}\bigg)\\
&-\half \sum\limits_{M=+,-} \big(D^\alpha\bar\Phi^M + i \bar\Phi^M \Gamma^\alpha\big) \big(D_\alpha\Phi^M - i \Gamma_\alpha\Phi^M\big)  - m_0 \left[(\bar\Phi^+ \Phi^+) - (\bar\Phi^- \Phi^-)\right] \\
&- \frac{\pi}{\kappa} (\bar\Phi^+ \Phi^+) (\bar\Phi^+ \Phi^+) - \frac{\pi}{\kappa} (\bar\Phi^- \Phi^-) (\bar\Phi^- \Phi^-)+ \frac{4\pi}{\kappa} (\bar\Phi^+ \Phi^+) (\bar\Phi^- \Phi^-) + \frac{2\pi}{\kappa} (\bar\Phi^+ \Phi^-) (\bar\Phi^- \Phi^+)\bigg],
\end{split}
\end{equation}
\end{widetext}
where $\Gamma^\alpha$ is the $U(N)$ gauge superfield. Note that there is no 
raising or lowering for the $SO(2)_R$ indices. 

In the rest of the section, we use the standard superfield expansions and show that we recover the component action eq.\ \eqref{eu3cs} from eq.\ \eqref{n3sup}. The superfield expansions in the Wess-Zumino gauge are given by
\begin{equation}
\begin{split}
\label{comp1}
&\Phi^+ = \phi^+ + \theta \psi^+ - \theta^2 F^+ ,\\
&\Phi^- = \phi^- + \theta \psi^- - \theta^2 F^- ,\\
&\Gamma^\alpha =  i \theta^\beta A_\beta^{~\,\alpha} - 2 \theta^2 \lambda^\alpha.
\end{split}
\end{equation}
One needs to integrate out the auxiliary fields $F^{\pm}, \bar{F}^{\pm}$ and the gaugino $\lambda^\alpha$ by using their equations of motion. The equations of motion for $F^{\pm}$ are given by
\begin{equation}
\begin{split}
\label{feom}
&\bar{F}^+ = m_0 \bar\phi^+ - \frac{2\pi}{\kappa} (\bar\phi^+ \phi^-) \bar\phi^- - \frac{4\pi}{\kappa} (\bar\phi^- \phi^-) \bar\phi^+ \\
&\qquad+ \frac{2\pi}{\kappa} (\bar\phi^+ \phi^+) \bar\phi^+ ,\\
&\bar{F}^- = - m_0 \bar\phi^- - \frac{2\pi}{\kappa} (\bar\phi^- \phi^+) \bar\phi^+ - \frac{4\pi}{\kappa} (\bar\phi^+ \phi^+) \bar\phi^- \\
&\qquad+ \frac{2\pi}{\kappa} (\bar\phi^- \phi^-) \bar\phi^- .
\end{split}
\end{equation}
The gaugino equation of motion is
\begin{equation}
\begin{split}
\label{gaugeeom}
(\lambda_\alpha)_i^{~j} = \frac{2\pi}{\kappa} \Big[ &-i (\bar\phi^{+})^j (\psi^+_{\alpha})_i + i (\bar\psi^+_\alpha)^j (\phi^+)_i \\
&- i (\bar\phi^{-})^j (\psi^-_{\alpha})_i + i (\bar\psi^-_\alpha)^j (\phi^-)_i \Big].
\end{split}
\end{equation}
Eliminating the auxiliary fields gives us the component action
\begin{equation}
\begin{split}
\label{compons}
&\mathcal{S}^{\texttt{Euc}}_{CS} = \int d^3x \, \bigg[ \frac{i\kappa}{4\pi} \epsilon^{\mu\nu\rho} \, \text{Tr}\bigg( A_\mu \del_\nu A_\rho - \frac{2i}{3} A_\mu A_\nu A_\rho\bigg) \\
&\quad+ D^\mu \bar\phi^+ D_\mu \phi^+ + D^\mu \bar\phi^- D_\mu \phi^- - i \bar\psi^+ \slashed{D} \psi^+ - i \bar\psi^- \slashed{D} \psi^-\\
&\quad+ m_0^{~2} \left(\bar\phi^+ \phi^+ + \bar\phi^- \phi^-\right) + m_0 \left(\bar\psi^+ \psi^+ - \bar\psi^- \psi^- \right) \\
&\quad+ \frac{4\pi m_0}{\kappa} \left( \bar\phi^+ \phi^+ + \bar\phi^- \phi^-\right) \left( \bar\phi^+ \phi^+ - \bar\phi^- \phi^-\right)\\
&+\frac{4\pi^2}{\kappa^2} \Big\{(\bar\phi^+ \phi^+) (\bar\phi^+ \phi^+) (\bar\phi^+ \phi^+) + (\bar\phi^- \phi^-) (\bar\phi^- \phi^-) (\bar\phi^- \phi^-) \\
&\quad+ 3 (\bar\phi^+ \phi^+) (\bar\phi^+ \phi^-) (\bar\phi^- \phi^+) + 3(\bar\phi^- \phi^-) (\bar\phi^+ \phi^-) (\bar\phi^- \phi^+)\Big\}\\
&+ \frac{2\pi}{\kappa} \Big\{ (\bar\phi^- \psi^-) (\bar\phi^- \psi^-) + (\bar\psi^- \phi^-) (\bar\psi^- \phi^-) + (\bar\phi^+ \psi^+) (\bar\phi^+ \psi^+) \\
&\quad+ (\bar\psi^+ \phi^+) (\bar\psi^+ \phi^+) -(\bar\phi^+ \psi^-) (\bar\psi^- \phi^+) -(\bar\psi^+ \phi^-) (\bar\phi^- \psi^+) \\
&\quad+ (\bar\phi^- \psi^-) (\bar\psi^- \phi^-) + (\bar\phi^+ \psi^+) (\bar\psi^+ \phi^+)\Big\}\\
&-\frac{4\pi}{\kappa} \Big\{ (\bar\phi^+ \psi^+) (\bar\phi^- \psi^-) + (\bar\phi^+ \psi^+) (\bar\psi^- \phi^-) +  (\bar\psi^+ \phi^+) (\bar\phi^- \psi^-) \\
&\quad+ (\bar\psi^+ \phi^+) (\bar\psi^- \phi^-) + (\bar\phi^+ \phi^+)(\bar\psi^- \psi^-) + (\bar\phi^- \phi^-)(\bar\psi^+ \psi^+) \\
&\quad+ (\bar\phi^+ \phi^-)(\bar\psi^- \psi^+) + (\bar\phi^- \phi^+)(\bar\psi^+ \psi^-) \Big\}\bigg].
\end{split}
\end{equation}
Identifying the $SU(2)_R$ doublets carefully as listed in \S \ref{N3con} we recover eq.\ \eqref{eu3cs}. Thus eq.\ \eqref{n3sup} is the correct representation of the mass-deformed $\mathcal{N} = 3$ Chern-Simons-Matter action in $\mathcal{N} = 1$ superspace.

\section{Concluding Comments}
\label{discuss}
In this paper, we presented the explicit form of the action and supersymmetry transformations for the mass-deformed $\mathcal{N} = 3$ supersymmetric $U(N)$ Chern-Simons theory coupled to fundamental matter. We also wrote down the mass-deformed action in  $\mathcal{N} = 1$ superspace, eq.\ \eqref{n3sup}. Our results fill a gap in the literature on Chern-Simons theories. We make use of the results presented in this paper in a follow up work \cite{InbasekarN3} to compute exact scattering amplitudes to all loop orders for the mass-deformed $\mathcal{N} = 3$ theory. 

As was mentioned in the introduction \S\ref{Intro}, there are several interesting straightforward directions to pursue following this paper. Firstly, one could further test the bosonization duality in the $\mN=3$ theory and compute correlation functions of spin-zero supercurrents following \cite{Inbasekar:2019wdw}, and beta functions to study RG flows and fixed points following \cite{Aharony:2019mbc}. The $2\to 2$ amplitudes computed in the $\mN=3$ theory in \cite{InbasekarN3} are tree level exact, and we expect that the amplitudes would be invariant under dual superconformal symmetry and Yangian symmetry \cite{Inbasekar:2017sqp,Yangian}. Lastly, it should be possible to compute arbitrary $n$-point amplitudes via BCFW recursion relations following \cite{Inbasekar:2017ieo}. We hope to report on some of these directions in the future.

\acknowledgments
We would like to thank Sachin Jain, Shiraz Minwalla and V. Vishal for collaboration in the initial stages of this project. We would also like to thank Ken Intriligator, Sachin Jain, Shiraz Minwalla, Naveen Prabhakar, Tarun Sharma and V. Umesh for helpful discussions. The work of KI is supported in part by a Center of Excellence supported by the Israel Science Foundation (grant number 1989/14), the US-Israel bi-national fund (BSF) grant number 2012383, and the Germany Israel bi-national fund GIF grant number I-244-303.7-2013 at Tel Aviv University and BSF grant number 2014707 at Ben Gurion University. The work of LJ is supported in part by the Infosys Endowment for the Study of the Quantum Structure of Spacetime. The work of AS is supported by the NSERC of Canada.

\appendix
\section{Notations and Conventions}
\label{conventions}
In this appendix, we setup the relevant notations and conventions. We work with the metric signature $(-,+,+)$.  We follow appendix A of \cite{Inbasekar:2015tsa} for the spinors, Dirac matrices and $\mN=1$ superspace conventions. Additional conventions in $\mN=1$ superspace are summarized in \cite{Gates:1983nr}. 

\subsection{Conventions for $SU(2)_R$ and $SO(2)_R$}\label{N3con}
We group the $SO(2)_R$ fields that appear in the action eq.\ \eqref{compons} into $SU(2)_R$ doublets as shown below,
\begin{eqnarray}\label{relations}
\phi^A = &\left(
\begin{array}{c} 
\phi^+\\
\phi^-
 \end{array}
\right)
\begin{array}{c} 
+ \\
-
 \end{array}
\ , \
\phi_A = &\left(
\begin{array}{c} 
-\phi^-\\
\phi^+
 \end{array}
\right)
\begin{array}{c} 
- \\
+
 \end{array}\nn\\
 \bph_A = &\left(
\begin{array}{c} 
\bph^+\\
\bph^-
 \end{array}
\right)
\begin{array}{c} 
- \\
+
 \end{array}\ , \
\bph^A = &\left(
\begin{array}{c} 
-\bph^-\\
\bph^+
 \end{array}
\right)
\begin{array}{c} 
+ \\
-
 \end{array}
\ , \
\end{eqnarray}
and
\begin{eqnarray}
\psi^A = \left(
\begin{array}{c} 
\psi^+\\
-\psi^-
 \end{array}
\right)
\begin{array}{c} 
+ \\
-
 \end{array}
\ , \
\psi_A = \left(
\begin{array}{c} 
\psi^-\\
\psi^+
 \end{array}
\right)
\begin{array}{c} 
- \\
+
 \end{array}\nn\\
\bps_A = \left(
\begin{array}{c} 
\bps^+\\
-\bps^-
 \end{array}
\right)
\begin{array}{c} 
- \\
+
 \end{array}
\ , \
\bps^A = \left(
\begin{array}{c} 
\bps^-\\
\bps^+
 \end{array}
\right)
\begin{array}{c} 
+\\
-
 \end{array}
.
\end{eqnarray}
The raising and lowering of the $R$-symmetry index is done via
\beq
\ph_A=\ph^B\ep_{BA}\ , \ \ps^A=\ep^{AB}\ps_B,
\eeq
with the convention $\ep_{12}=\ep^{12}=1$.  This implies the rule
\beq
\ep_{AB}\ep^{BC}=-\de_A^{ \ C} .
\eeq
Note that because of this the mixed identity tensors are antisymmetric.

The matrix $C_{\alpha\beta}$ is the antisymmetric charge-conjugation matrix which is used to raise and lower spinor indices,
\begin{equation}
\label{spinsymb}
C_{\alpha\beta} = -\, C_{\beta\alpha} = 
{\begin{pmatrix}
0 & -i \\
i & 0
\end{pmatrix}} =  -\, C^{\alpha\beta},
\end{equation}
with the raising/lowering defined via $\psi^{\alpha} = C^{\alpha\beta} \psi_{\beta},
\psi_{\alpha} = \psi^{\beta} C_{\beta\alpha}$.

The fundamental representation $\mathbf{2}$ of $SU(2)$ is pseudo-real and it acts on
complex doublets \cite{Polchinski:1998rr}. The $*$ acts on $SU(2)$ spinors via charge
conjugation. For a bosonic $SU(2)$ matrix $M$ charge conjugation is defined as
\beq
(M)^*= \si^2 \bar{M} \si^2\ ,
\eeq
This implies that the complex conjugate representation is related to the original one by a
similarity transformation. As a result, for instance
\beq
((\si_3)_A^{\ B})^*= (\bar{\si_3})^A_{\ B} = -(\si_3)_B^{ \ A}.
\eeq
Thus for the charge conjugate version of a matrix $M_A^{ \ B}$, the index structure is understood to
be $\bar{M}^A_{\ B}=-M_B^{\ A}$. We define our complex conjugated fields as 
\beq
\bar{\ph}^A= (\ph_A)^* \ , \bph_A=(\ph^A)^*.
\eeq
The rule for complex conjugation then induces the rule
\beq
\bph^A = \bph_B \ep^{BA} \ , \bph_A = \ep_{AB}\bph^B,
\eeq
from which it follows that
\beq
\bph^A\ph_A= \bph_B \ep^{BA}\ph_A=\bph_B\ph^B,
\eeq
and 
\beq
(\ep_{AB})^*=\ep^{AB}.
\eeq

Note that the product of two pseudo-real representations is real. This induces the
rule $Q_{DE}=\ep_{DB}\ep_{EC}(Q_{BC})^*$. It follows that
$$Q_{DE}=\ep_{DB}\ep_{EC}\bar{Q}^{BC} =\bar{Q}_{DE},$$
i.e. the supercharges are real.

\section{Supersymmetric Invariance of the Free $\mathcal{N} = 3$ Wess-Zumino Model}
\label{massdeform}
In this appendix, we detail the intermediate steps involved in the computation of the supersymmetric variation of the action $S_{WZ}^{free}$, given in eq.\eqref{actn3}. We determine the conditions that the coefficients $(\chi_i, \tilde\chi_i)$, defined in  eq.\eqref{gensusy3},  must satisfy so that the action is  invariant under $\mN=3$ supersymmetry, eq.\eqref{invcond1}. Using the supersymmetry transformations for the scalar and spinor fields given in eq.\eqref{gensusy3}, it is straightforward to calculate the action of the supercharge $Q_{BC\alpha}$ on $S_{WZ}^{free}$. For instance, the variation of the first term in the action eq.\eqref{actn3} produces
\begin{equation}
\begin{split}
\label{varn1}
&Q_{BC\alpha} \bigg(- \int d^3x \, \del^\mu \bar\phi^A \, \del_\mu \phi_A\bigg) = \\
&- \int d^3x \Big[ \tilde\chi_1 \, \del^\mu \bar\psi_{\alpha B} \, \del_\mu\phi_C + \tilde\chi_1 \, \del^\mu \bar\psi_{\alpha C} \, \del_\mu\phi_B \\
&\qquad\quad+ \chi_1 \, \del^\mu \bar\phi_B \, \del_\mu\psi_{\alpha C} +  \chi_1 \, \del^\mu \bar\phi_C \, \del_\mu\psi_{\alpha B} \Big],
\end{split}
\end{equation}
where the supercharge $Q_{BC\alpha}$ acts distributively on the products of field operators in the LHS. The supersymmetric variation of the second term in the action eq.\eqref{actn3} gives
\begin{equation}
\begin{split}
\label{varn2}
&Q_{BC\alpha} \left(- \int d^3x \, \bar\phi^A M_A^{~\,D} M_D^{~\,E} \phi_E\right) = \\
&- \int d^3x \Big[ \tilde\chi_1 \bar\psi_{\alpha B} M_C^{~\,D} M_D^{~\,E}  \phi_E + \tilde\chi_1 \bar\psi_{\alpha C} M_B^{~\,D} M_D^{~\,E}  \phi_E \\
&\qquad\quad- \chi_1 \bar\phi^A M_A^{~\,D} M_{DC} \, \psi_{\alpha B} - \chi_1 \bar\phi^A M_A^{~\,D} M_{DB} \, \psi_{\alpha C} \Big].
\end{split}
\end{equation}
Similarly, from the third term in the action, we get the contribution\footnote{The distributive action of the supercharge $Q_{AB\alpha}$ on a product of fields $\xi_1, \xi_2, \xi_3 \ldots$ is given by
\begin{equation*}
Q_{AB\alpha} (\xi_1 \xi_2 \xi_3\ldots) = (Q_{AB\alpha} \xi_1) \xi_2 \xi_3\ldots + \xi_1 (Q_{AB\alpha} \xi_2 ) \xi_3 \ldots + \ldots
\end{equation*}
Note that the above distributive property holds true if $\xi_1, \xi_2, \xi_3 \ldots$ are bosonic in nature. Since the supercharge is itself fermionic, every time it crosses another fermionic field/operator, an additional `$-$' sign needs to be inserted in the equation above.}
\begin{equation}
\begin{split}
\label{varn3}
&Q_{BC\alpha} \bigg(i \int d^3x\, \bar\psi^{\beta A} \del_{\beta}^{~\,\gamma} \psi_{\gamma A}\bigg) = \\
&i \int d^3x \Big[ \tilde\chi_2 \, \del_\alpha^{~\,\beta} \bar\phi_B \, \del_\beta^{~\,\gamma} \psi_{\gamma C} + \tilde\chi_2 \, \del_\alpha^{~\,\beta} \bar\phi_C \, \del_\beta^{~\,\gamma} \psi_{\gamma B} \\
&\qquad\quad+ \tilde\chi_3\, \bar\phi_B M_C^{~\,A} \del_{\alpha}^{~\, \gamma} \psi_{\gamma A} + \tilde\chi_3\, \bar\phi_C M_B^{~\,A} \del_{\alpha}^{~\, \gamma} \psi_{\gamma A}  \\
&\qquad\quad- \chi_2\, \bar\psi^\beta_B \del_{\beta}^{~\,\gamma} \del_{\gamma\alpha} \phi_C - \chi_2\, \bar\psi^\beta_C \del_{\beta}^{~\,\gamma} \del_{\gamma\alpha} \phi_B \\
&\qquad+ \chi_3 \, \bar\psi^{\beta A} \del_{\alpha\beta} \phi_B M_{CA} + \chi_3 \, \bar\psi^{\beta A} \del_{\alpha\beta} \phi_C M_{BA} \Big].
\end{split}
\end{equation}
Finally, the variation of the fourth term in the action produces
\begin{equation}
\begin{split}
\label{varn4}
&Q_{BC\alpha} \bigg( \int d^3x\,  \bar\psi^{\beta A} M_A^{~\,D} \psi_{\beta D} \bigg) \\
&= \int d^3x \Big[ \tilde\chi_2 \, \del_\alpha^{~\,\beta} \bar\phi_B M_C^{~\,D} \psi_{\beta D} + \tilde\chi_2 \, \del_\alpha^{~\,\beta} \bar\phi_C M_B^{~\,D} \psi_{\beta D}\\
&\qquad+ \tilde\chi_3\, \bar\phi_B M_{C}^{~\,A} M_A^{~\,D} \psi_{\alpha D} + \tilde\chi_3\, \bar\phi_C M_{B}^{~\,A} M_A^{~\,D} \psi_{\alpha D} \\
&\qquad+ \chi_2 \, \bar\psi^{\beta A} M_{AC} \del_{\alpha\beta} \phi_B +\chi_2 \, \bar\psi^{\beta A} M_{AB} \del_{\alpha\beta} \phi_C \\
&\qquad+ \chi_3\, \bar\psi^A_\alpha M_A^{~\,D} \phi_B M_{CD} + \chi_3\, \bar\psi^A_\alpha M_A^{~\,D} \phi_C M_{BD} \Big].
\end{split}
\end{equation}

For the supersymmetric invariance of the action $S_{WZ}^{free}$, the sum of the RHS of eqs.\eqref{varn1}-\eqref{varn4} must vanish, upto surface terms. As discussed in section \ref{mdn3}, this is possible for two distinct choices of the mass matrix $M_A^{~\,B}$. The requirement of supersymmetric invariance of the action puts constraints on the coefficients $\chi$ for both the cases. In the first case, where we have $M_A^{~\,B} = m \delta_A^{~\,B}$, we get the set of conditions given in eq.\eqref{cond1} for the coefficients $(\chi_i, \tilde\chi_i)$. On the other hand, for the second case, where $M_A^{~\,B} = m (\sigma_3)_A^{~\,B}$, one finds the set of conditions given in eq.\eqref{cond2}.

\section{Supersymmetric Invariance of the Lorentzian Massless $\mathcal{N} = 3$ Chern-Simons-Matter Theory}
\label{varn3l}
In this appendix, we verify that the action eq.\eqref{lo3cs} is invariant under $\mathcal{N} = 3$ supersymmetry. The variation is done by acting the supercharge $Q_{XY\alpha}$ on the action $S_{CS}$. To start, based on dimensional analysis and Lorentz invariance, we write the general supersymmetry transformations for the $\mathcal{N}=3$ theory as 
\begin{equation}
\begin{split}
\label{sustn3}
&Q_{BC\alpha} \phi_A = \chi_1 \, \psi_{\alpha (B} \, \epsilon_{C)A} ,\\
&Q_{BC\alpha} \bar\phi^A = - \chi_1^* \, \bar\psi_{\alpha (B} \, \delta_{C)}^{~~A} ,\\
&Q_{BC\alpha} \psi_{\beta A} = -i \chi_1^* \, D_{\alpha\beta} \phi_{(B} \, \epsilon_{C)A} + \chi_2 \, C_{\alpha\beta} (\bar\phi_A \phi_{(B}) \phi_{C)} \\
&\qquad\qquad+ \chi_3 \, C_{\alpha\beta} (\bar\phi_{(B} \phi_{A}) \phi_{C)} + \chi_5 \, C_{\alpha\beta} (\bar\phi_{(B} \phi_{C)}) \phi_{A},\\
&Q_{BC\alpha} \bar\psi^{\beta A} = i \chi_1 \, D_{\alpha}^{~\,\beta} \bar\phi_{(B} \, \delta_{C)}^{~~A} + \tilde\chi_2 \, \delta_{\alpha}^{~\,\beta}(\bar\phi_{(B} \phi^{A}) \bar\phi_{C)} \\
&\qquad\qquad+ \tilde\chi_3 \, \delta_{\alpha}^{~\,\beta} (\bar\phi^A \phi_{(B}) \bar\phi_{C)} + \tilde\chi_5 \, \delta_{\alpha}^{~\,\beta} (\bar\phi_{(B} \phi_{C)}) \bar\phi^{A},\\
&Q_{BC\alpha} A^a_\mu = \chi_4 (\gamma_\mu)_\alpha^{~\,\beta} \bar\phi^i_{(B} (T^a)_i^{~j} \psi_{C)\beta j} \\
&\qquad\qquad+ \tilde\chi_4 (\gamma_\mu)_\alpha^{~\,\beta} \bar\psi^i_{\beta (B} (T^a)_i^{~j} \phi_{C)j} .
\end{split}
\end{equation} 
The invariance of the action under the supersymmetry transformations imposes conditions on the unknown coefficients $(\chi_i, \tilde\chi_i)$, an idea we have already used in appendix \ref{massdeform} to fix the action for the mass-deformed free $\mathcal{N} = 3$ Wess-Zumino model. For instance, making use of the supersymmetry transformations in eq.\eqref{sustn3}, the variation of the first term of the action eq.\eqref{lo3cs} gives\footnote{Note that we can rewrite the supersymmetry transformation of the gauge field $A_\mu$ in eq.\eqref{sustn3} as
\begin{equation*}
\label{susyg}
Q_{BC\alpha} (A_\mu)_i^{~j} = \half \Big[ \chi_4 (\gamma_\mu)_\alpha^{~\,\beta} \bar\phi^j_{(B} \psi_{C)\beta i} + \tilde\chi_4 (\gamma_\mu)_\alpha^{~\,\beta} \bar\psi^j_{\beta (B} \phi_{C)i}\Big],
\end{equation*}
where we have made use of the identity
\begin{equation*}
\label{git}
(T^a)_i^{~j} (T^a)_k^{~l} = \half \, \delta_{i}^{~l} \delta_{k}^{~j}.
\end{equation*}}
\begin{equation}
\begin{split}
\label{xy1}
Q_{XY\alpha} \bigg( &\int d^3x \, \text{Tr}\bigg(\frac{\kappa}{4\pi} \, \epsilon^{\mu\nu\rho} A_\mu \del_\nu A_\rho\bigg) \bigg) = \\
&\frac{\kappa}{4\pi} \, \epsilon^{\mu\nu\rho} \int d^3x \,\Big[ \chi_4 (\gamma_\mu)_\alpha^{~\,\beta} \bar\phi_{(X} \, \del_\nu A_\rho\,  \psi_{Y)\beta} \\
&\qquad\qquad\qquad+ \tilde\chi_4 (\gamma_\mu)_\alpha^{~\,\beta} \bar\psi_{\beta (X} \, \del_\nu A_\rho \, \phi_{Y)}\Big].
\end{split}
\end{equation}

One can similarly compute the supersymmetric variation of the rest of the terms in the action eq.\eqref{lo3cs}. These terms should mutually cancel, upto surface terms, to make the action supersymmetry invariant. The mutual cancellation of terms containing one boson and three fermion fields gives the conditions
\begin{equation}
\begin{split}
\label{condn31}
&\chi_1^* = \chi_1,\\
&\chi_4 = \tilde\chi_4 = - \frac{4\pi}{\kappa} \chi_1,
\end{split}
\end{equation}
with $\chi_1$ being any real number. In deriving eq.\eqref{condn31}, we have made use of the Fierz identity
\begin{equation}
\chi_\alpha \xi^\beta \eta_\beta + \xi_\alpha \chi^\beta \eta_\beta + \eta_\alpha \xi^\beta \chi_\beta = 0,
\end{equation}
which shuffles spinor indices on products of anti-commuting fields, along with the Fierz identities for the rearrangement of $R$-symmetry indices, given by
\begin{align}
\label{fierz2}
&\chi_A \bar\xi^B \bar\eta_B + \bar\xi_A \chi^B \bar\eta_B + \bar\eta_A \bar\xi^B \chi_B = 0,\\
\label{fierz3}
&\bar\chi_A \xi^B \eta_B + \xi_A \bar\chi^B \eta_B + \eta_A \xi^B \bar\chi_B = 0,
\end{align}
where $\chi, \eta$ are mutually anti-commuting fields, whereas $\xi$ commutes. Next, we collect all the terms which contain a single gauge field contribution, but do not include derivatives. The mutual cancellation of such terms gives
\begin{equation}
\begin{split}
\label{condn32}
&\chi_2 = \frac{2\pi}{\kappa} \chi_1, ~~\chi_3 = 0, ~~\chi_5 = \frac{2\pi}{\kappa} \chi_1 ,\\
&\tilde\chi_2 = \frac{2\pi}{\kappa} \chi_1, ~~\tilde\chi_3 = 0, ~~\tilde\chi_5 = -\frac{2\pi}{\kappa} \chi_1 .
\end{split}
\end{equation}
Once again, in deriving eq.\eqref{condn32}, we have made use of Fierz rearrangement identities in $R$-symmetry space, given by
\begin{align}
\label{fierz4}
&\chi_A \bar\xi^B \bar\eta_B + \bar\xi_A \chi^B \bar\eta_B - \bar\eta_A \bar\xi^B \chi_B = 0, \\
\label{fierz5}
&\chi_A \xi^B \bar\eta_B - \xi_A \chi^B \bar\eta_B - \bar\eta_A \xi^B \chi_B = 0,\\
\label{fierz6}
&\bar\chi_A \xi^B \bar\eta_B + \xi_A \bar\chi^B \bar\eta_B - \bar\eta_A \xi^B \bar\chi_B = 0,\\
\label{fierz7}
&\bar\chi_A \xi^B \eta_B + \xi_A \bar\chi^B \eta_B - \eta_A \xi^B \bar\chi_B = 0,
\end{align}
where $\chi, \xi, \eta$ are all mutually commuting fields. Eqs.\eqref{condn31} and \eqref{condn32} completely fix the supersymmetry transformations up to one arbitrary numerical factor $\chi_1$. It is straight forward to check that with the choice of $(\chi_i, \tilde\chi_i)$ we have made, eqs. \eqref{condn31} and \eqref{condn32}, the remaining terms in the variation of the $\mathcal{N} = 3$ action also vanish.

\bibliography{refer}
\onecolumngrid
\end{document}